\newcommand{\openone}{\leavevmode\hbox{\small1\normalsize\kern-.33em1}}
\newcommand{\matriz}[1]{{\mathbf{#1}}}  
\newcommand{\coefr}{R}  
\newcommand{\coeft}{T}  
\newcommand{\Tr}{\mathop{\mathrm{Tr}}\nolimits}
\newcommand{\re}{\mathop{\mathrm{Re}}\nolimits}
\newcommand{\im}{\mathop{\mathrm{Im}}\nolimits}
\newcommand{\Par}{\mathcal{P}}
\newcommand{\T}{\mathcal{T}}
\newcommand{\PT}{\mathcal{P}\mathcal{T}}

\documentclass[symmetry,article,accept,moreauthors,12pt,a4paper]{mdpi}

\lastpage{x}
\doinum{10.3390/------}
\pubvolume{6}
\pubyear{2014}
\history{Received: }

\usepackage{amsmath,amssymb,graphicx,bm,color,hyperref}
\usepackage{soul}
\Title{Invisibility and $\PT$ Symmetry: A Simple Geometrical Viewpoint}

\Author{Luis L. S\'anchez-Soto *  and Juan J. Monz\'on}

\address[1]{Departamento de \'Optica,  Facultad de F\'{\i}sica,
Universidad Complutense, 28040~Madrid,  Spain; 
\linebreak E-Mail: jjmonzon@opt.ucm.es}

\corres{E-Mail: lsanchez@fis.ucm.es; 
 \linebreak Tel.: +34 913944680; Fax: +34 913944683.}

\abstract{We give a simplified account of the properties of the
  transfer matrix for a complex one-dimensional potential,
  paying special attention to the particular instance of
  unidirectional invisibility. In appropriate variables, invisible
  potentials appear as performing null rotations, which lead to the
  helicity-gauge symmetry of massless particles. In hyperbolic
  geometry, this can be interpreted, via M\"{o}bius transformations,
  as parallel displacements, a geometric action that has no Euclidean
  analogy.}


\begin{document}

\vspace{-12pt}
\section{Introduction}

The work of Bender and coworkers~\cite{Bender:1998fk,Bender:1999vn,
  Bender:2002uq,Bender:2003zr,Bender:2007ve,Bender:2010ly} has
triggered considerable efforts to understand complex potentials that
have neither parity ($\Par$) nor time-reversal symmetry ($\T$), yet
they retain combined $\PT$ invariance.  These systems can exhibit real
energy eigenvalues, thus suggesting a plausible generalization of
quantum mechanics. This speculative concept has motivated an ongoing
debate in several forefronts~\cite{Assis:2010fk,Moiseyev:2011qy}.

Quite recently, the prospect of realizing $\PT$-symmetric potentials
within the framework of optics has been
put forward~\cite{El-Ganainy:2007mb,Bendix:2009gb} and experimentally
tested~\cite{Ruter:2010ss}.  The complex refractive index takes on
here the role of the potential, so they can be realized through a
judicious inclusion of index guiding and gain/loss regions. These
$\PT$-synthetic materials can exhibit several intriguing
features~\cite{Makris:2008jl,Longhi:2009uq,Sukhorukov:2010ys},
one of which will be the main interest of this paper, namely,
unidirectional  invisibility~\cite{Ahmed:2005fk,Lin:2011zr,Longhi:2011ve}.

In all these matters, the time-honored transfer-matrix method is
particularly germane~\cite{Sanchez-Soto:2012bh}. However, a quick look
at the literature immediately reveals the different backgrounds and
habits in which the transfer matrix is used and the very little cross
talk between them.

To remedy this flaw, we have been capitalizing on a number of
geometrical concepts to gain further insights into the behavior of
one-dimensional scattering~\cite{Monzon:1999eh,Monzon:1999fk,
  Monzon:2001b,Yonte:2002,Monzon:2002mz,Barriuso:2003,
  Barriuso:2004,Barriuso:2009}. Indeed, when one think in a unifying
mathematical scenario, geometry immediately comes to mind. Here, we
keep going this program and examine the action of the transfer
matrices associated to invisible scatterers. Interestingly enough,
when viewed in SO(1, 3), they turn to be nothing but parabolic Lorentz
transformations, also called null rotations, which play a crucial role
in the determination of the little group of massless particles.
Furthermore, borrowing elementary techniques of hyperbolic geometry,
we reinterpret these matrices as parallel displacements, which are
motions without Euclidean counterpart.

We stress that our formulation does not offer any inherent advantage
in terms of efficiency in solving practical problems; rather, it
furnishes a general and unifying setting to analyze the transfer
matrix for complex potentials, which, in our opinion, is more than a
curiosity.

\section{Basic Concepts on Transfer Matrix}

To be as self-contained as possible, we first briefly review some
basic facts on the quantum scattering of a particle of mass $m$ by a
local complex potential $V(x)$ defined on the real line
$\mathbb{R}$~\cite{Muga:2004ys,Levai:2000kx,Ahmed:2001ys,
 Ahmed:2001dz,Mostafazadeh:2009ve,Cannata:2007kx,Chong:2011vn,
Ahmed:2012qf}. Although much of the renewed interest in this topic has
been fuelled by the remarkable case of $\PT$ symmetry, we do not use this
extra assumption in this Section.

The problem at hand is governed by the time-independent
Schr\"{o}dinger equation
\begin{equation}
  \label{eq:Schrind}
  H \Psi( x)  =
  \left [ - \frac{d^2}{dx^2} + U(x) \right ] \Psi (x) =
  \varepsilon \, \Psi (x) \,
\end{equation}
where $\varepsilon = 2 m E/\hbar^2$ and $U(x) = 2 mV(x)/\hbar^2$, $E$
being the energy of the particle.  We assume that $U(x) \rightarrow 0$
fast enough as $x \rightarrow \pm \infty$, although the treatment can
be adapted, with minor modifications, to cope  with potentials for
which the limits $U_{\pm} = \lim_{x \rightarrow \pm \infty} U(x)$ are
different.

Since $U(x)$ decays rapidly as $|x| \rightarrow \infty$, solutions of
\eqref{eq:Schrind} have the asymptotic behavior
\begin{equation}
  \label{eq:movers}
  \Psi(x) = \left \{
    \begin{array}{ll}
      A_{+} e^{+ i k x} + A_{-} e^{- i k x}
      \quad & x \rightarrow  - \infty   \\
      B_{+} e^{+ i k x} + B_{-} e^{- i k x}
      & x \rightarrow \infty  \\
    \end{array}
  \right .
\end{equation}
Here, $k^{2} = \varepsilon$, $A_{\pm}$ and $B_{\pm}$ are $k$-dependent
complex coefficients (unspecified, at this stage), and the subscripts
$+$ and $-$ distinguish right-moving modes $\exp (+ i kx)$ from
left-moving modes $\exp(- i kx)$, respectively.

The  problem requires to work out the exact solution of (\ref{eq:Schrind}) and
invoke the appropriate boundary conditions, involving not only the
continuity of $\Psi(x)$ itself, but also of its derivative.  In this
way, one has two linear relations among the coefficients $A_{\pm}$ and
$B_{\pm}$, which can be solved for any amplitude pair in terms of the
other two; the result can be expressed as a matrix equation that
translates the linearity of the problem.  Frequently, it is more
advantageous to specify a linear relation between the wave amplitudes
on both sides of the scatterer, namely,
\begin{equation}
  \label{M}
  \left (
    \begin{array}{c}
     B_{+} \\
     B_{-}
    \end{array}
  \right ) =
  \matriz{M}
  \left (
    \begin{array}{c}
      A_{+} \\
      A_{-}
    \end{array}
  \right ) \,
\end{equation}
$\matriz{M}$ is the transfer matrix, which depends in a complicated
way on the potential $U (x)$. Yet one can extract a good deal of information
without explicitly calculating it: let us apply~(\ref{M})
successively to a right-moving [$(A_{+} =1, B_{-} = 0)$] and to a
left-moving wave [$(A_{+} =0, B_{-} = 1)$], both of unit
amplitude. The result can be displayed as
\begin{equation}
  \label{eq:resu}
  \left (
    \begin{array}{c}
      \coeft^{\ell} \\
      0
    \end{array}
  \right )
  =
  \matriz{M}
  \left (
    \begin{array}{c}
      1 \\
      \coefr^{\ell}
    \end{array}
  \right ) \, ,
  \qquad
  \left (
    \begin{array}{c}
      \coefr^{r} \\
      1
    \end{array}
  \right )
  =
  \matriz{M}
  \left (
    \begin{array}{c}
      0 \\
      \coeft^{r}
    \end{array}
   \right ) \,
\end{equation}
where $\coeft^{\ell, r}$ and $\coefr^{\ell, r}$ are the transmission
and reflection coefficients for a wave incoming at the potential from
the left and from the right, respectively, defined in the standard way
as the quotients of the pertinent fluxes~\cite{Boonserm:2010fk}.

With this in  mind, Equation~\eqref{eq:resu} can be thought of as a linear
superposition of the two independent solutions
\begin{equation}
  \label{eq:asym}
  \Psi_{k}^{\ell} (x ) = \left \{
    \begin{array}{lll}
      e^{+ i k x}  +  \coefr^{\ell} (k) \, e^{- i k x}
      \quad &  x \rightarrow  - \infty \, , \\
      \qquad \quad  \;\, \coeft^{\ell} (k) \, e^{+  i k x}
      & x \rightarrow  \infty \, , \\
    \end{array}
  \right .
  \quad
  \Psi_{k}^{r} (x ) = \left \{
    \begin{array}{lll}
      \qquad \quad \; \, \coeft^{r} (k) \, e^{- i k x}
      & x \rightarrow  - \infty \,  \\
      e^{- i k x}  + \coefr^{r} (k) \, e^{+ i k x}
      \quad &  x \rightarrow  \infty  \,
    \end{array}
  \right .
\end{equation}
which is consistent with the fact that, since $\varepsilon >0$, the
spectrum of the Hamiltonian (\ref{eq:Schrind}) is continuous and there
are two linearly independent solutions for a given value of
$\varepsilon$.  The wave function $\Psi_{k}^{\ell}
(x)$ represents a wave incident from $- \infty$ [$\exp (+ i k
x)$] and the interaction with the potential produces a reflected wave
[$\coefr^{\ell} (k) \exp(- i k x)$] that escapes to $- \infty$ and a
transmitted wave [$\coeft^{\ell} (k) \exp( + i k x)$] that moves off to
$+ \infty$. The solution $\Psi_{k}^{r} (x)$ can be interpreted in a
similar fashion.

Because of the Wronskian of the solutions \eqref{eq:asym} is
independent of $x$, we can compute \linebreak $W (\Psi_{k}^{\ell}, \Psi_{k}^{r} )
= \Psi_{k}^{\ell} \Psi_{k}^{r \, \prime} - \Psi_{k}^{\ell \, \prime}
\Psi_{k}^{r}$ first for $x \rightarrow - \infty$ and then for $x
\rightarrow \infty$; this gives
\begin{equation}
  \label{eq:Wron}
  \frac{i}{2k} W ( \Psi_{k}^{\ell}, \Psi_{k}^{r} )  =
  \coeft^{r} (k) = \coeft^{\ell} (k) \equiv \coeft (k) \,
\end{equation}
We thus arrive at the important conclusion that, irrespective of the
potential, the transmission coefficient is always independent of the
input direction.

Taking this constraint into account, we go back to the system
(\ref{eq:resu}) and write the solution for $\matriz{M}$ as
\begin{equation}
  \label{eq:ent}
  M_{11} (k)= \displaystyle
  \coeft (k)  - \frac{\coefr^{\ell} (k) \coefr^{r}(k)}{\coeft (k)} \, ,
  \quad
  M_{12} (k) = \displaystyle
  \frac{\coefr^{r}(k)}{\coeft(k)} \, ,
  \quad
  M_{21} (k) =\displaystyle
  - \frac{\coefr^{\ell} (k)}{\coeft(k)} \,  ,
  \quad
  M_{22} (k)=
  \displaystyle\frac{1}{\coeft(k)} \,
\end{equation}
A straightforward check shows that $\det \matriz{M} = +1$, so $\matriz{M}
\in$ SL(2, $\mathbb{C})$; a result that can be drawn from
a number of alternative and more elaborate
arguments~\cite{Mostafazadeh:2009yo}.

One could also relate outgoing amplitudes to the incoming ones  (as
they are often the magnitudes one can externally control): this
is precisely the scattering matrix, which can be concisely formulated as
\begin{equation}
  \label{S}
  \left (
    \begin{array}{c}
      B_{+} \\
      A_{-}
    \end{array}
  \right ) =
  \matriz{S}
  \left (
    \begin{array}{c}
      A_{+} \\
      B_{-}
    \end{array}
  \right ) \,
\end{equation}
with matrix elements
\begin{equation}
  \label{eq:SM}
  S_{11} (k) = \coeft(k) \, ,
  \qquad
  S_{12} (k) = \coefr^{r} (k) \, ,
  \qquad
  S_{21} (k) =  \coefr^{\ell} (k) \,  ,
  \qquad
  S_{22} (k) =  \coeft (k) \,
\end{equation}

Finally, we stress that transfer matrices are very convenient
mathematical objects.  Suppose that $V_{1}$ and $V_{2}$ are potentials
with finite support, vanishing outside a pair of adjacent intervals
$I_{1}$ and $I_{2}$. If $\matriz{M}_{1}$ and $\matriz{M}_{2}$ are the
corresponding transfer matrices, the total system (with support $I_{1}
\cup I_{2}$) is described by
\begin{equation}
  \label{propag}
  \matriz{M} = \matriz{M}_{1}  \, \matriz{M}_{2} \,
\end{equation}
This property is rather helpful: we can connect simple scatterers to
create an intricate potential landscape and determine its transfer
matrix by simple multiplication. This is a common instance in optics,
where one routinely has to treat multilayer stacks. However, this important
property does not seem to carry over into the scattering matrix in any
simple way~\cite{Aktosun:1992kx,Aktosun:1996fj}, because the incoming
amplitudes for the overall system cannot be obtained in terms of the
incoming amplitudes for every subsystem.

\section{Spectral Singularities}

The scattering solutions \eqref{eq:asym} constitute quite an intuitive
way to attack the problem and they are widely employed in physical
applications. Nevertheless, it is sometimes advantageous to look at
the fundamental solutions of \eqref{eq:Schrind} in terms of left-
and right-moving modes, as we have already used in
\eqref{eq:movers}.

Indeed, the two independent solutions of \eqref{eq:Schrind} can be
formally written down as~\cite{Marchenko:1986fk}
\begin{eqnarray}
  \label{eq:scint}
  \Psi^{(+)}_{k} (x) & = & e^{+ i kx} + \int_{x}^{\infty} \!\! K_{+} (x, x^{\prime})
  e^{+ i  k x^{\prime}}  dx^{\prime} \,  \nonumber\\
  & & \\
  \Psi_{k}^{(-)} (x) & = & e^{- ikx} + \int_{-\infty}^{x} \!\! K_{-} (x, x^{\prime})
  e^{- i  k x^{\prime}}  dx^{\prime} \,  \nonumber
\end{eqnarray}
The kernels $K_{\pm} (x, x^{\prime})$ enjoy a number of interesting
properties.  What matters for our purposes is that the resulting
$\Psi_{k}^{(\pm)} (x)$ are analytic with respect to $k$ in
$\mathbb{C}_{+} = \{ z\in\mathbb{C} | \im z >0 \}$ and continuous on
the real axis. In addition, it is clear that
\begin{equation}
  \label{eq:Jostasym}
  \Psi^{(+)}_{k} (x) = e^{+ i k x}   \quad   x \rightarrow \infty \, ,
  \qquad
  \Psi^{(-)}_{k} (x) = e^{- i k x}  \quad   x \rightarrow - \infty \,
\end{equation}
that is, they are the Jost functions for this
problem~\cite{Mostafazadeh:2009ve}.

Let us look at the Wronskian of the Jost functions $W(\Psi_{k}^{(-)} ,
\Psi_{k}^{(+)} )$, which, as a function of $k$, is analytical in
$\mathbb{C}_{+}$. A spectral singularity is a point $k_{\ast} \in
\mathbb{R}_{+}$ of the continuous spectrum of the Hamiltonian~(\ref{eq:Schrind}) such that
\begin{equation}
  \label{eq:specsing}
  W (\Psi_{k_{\ast}}^{(-)} , \Psi_{k_{\ast}}^{(+)} ) = 0 \,
\end{equation}
so $\Psi_{k}^{(\pm)} (x)$ become linearly dependent at $k_{\ast}$ and
the Hamiltonian is not diagonalizable. In fact, the set of zeros of
the Wronskian is bounded, has at most a countable number of elements
and its limit points can lie in a bounded subinterval of the real
axis~\cite{Tunca:1999vl}. There is an extensive theory of spectral
singularities for~\eqref{eq:Schrind} that was started by
Naimark~\cite{Naimark:1960bv}; the interested reader is referred to,
e.g., Refs.~\cite{Pavlov:1967kb,Naimark:1968qr,Samsonov:2005qq,
Andrianov:2010ts,Chaos-Cador:2013rf} for further~details.

The asymptotic behavior of $\Psi_{k}^{\pm} (x)$ at the opposite
extremes of $\mathbb{R}$ with respect to those in \eqref{eq:Jostasym}
can be easily worked out by a simple application of the transfer
matrix (and its inverse); viz,
\begin{eqnarray}
  \label{eq:Jostopo}
  \Psi^{(-)}_{k} (x) & = & M_{12} e^{+ i k x} + M_{22} e^{-ikx}
  \quad   x \rightarrow \infty \,  \nonumber \\
  & & \\
  \Psi^{(+)}_{k} (x) & = & M_{22} e^{+ i k x} - M_{21} e^{-ikx}
  \quad   x \rightarrow - \infty \,  \nonumber
\end{eqnarray}
Using $\Psi_{k}^{\pm} (x)$ in  (\ref{eq:Jostasym}) and
(\ref{eq:Jostopo}),  we can calculate
\begin{equation}
  \label{eq:WrJost}
  \frac{i}{2k} W ( \Psi_{k}^{(-)}, \Psi_{k}^{(+)} ) = M_{22} (k) \,
\end{equation}
Upon comparing with the definition (\ref{eq:specsing}), we can
reinterpret the spectral singularities as the real zeros of
$M_{22}(k)$ and, as a result, the reflection and transmission
coefficients diverge therein. The converse holds because $M_{12} (k)$
and $M_{21}(k)$ are entire functions, lacking
singularities.  This means that, in an optical scenario,  spectral
singularities correspond to lasing
thresholds~\cite{Schomerus:2010dk,Longhi:2010kx,Mostafazadeh:2013uf}.

One could also consider the more general case that the
Hamiltonian~(\ref{eq:Schrind}) has, in addition to a continuous
spectrum corresponding to $k \in \mathbb{R}_{+}$, a possibly complex
discrete spectrum. The latter corresponds to the square-integrable
solutions of that represent bound states. They are also zeros of
$M_{22}(k)$, but unlike the zeros associated with the spectral
singularities these must have a positive imaginary
part~\cite{Mostafazadeh:2009yo}.

The eigenvalues of $\matriz{S}$ are
\begin{equation}
  \label{eq:2}
  s_{\pm} = \frac{1}{M_{22}(k) } \left [
    1 \pm \sqrt{1- M_{11} (k)  M_{22}(k)} \right ]  \,
\end{equation}
At a spectral singularity, $s_{+}$ diverges, while $s_{-} \rightarrow
M_{11} (k) /2$, which suggests identifying spectral singularities with
resonances with a vanishing width.

\section{Invisibility and $\PT$ Symmetry}

As heralded in the Introduction, unidirectional invisibility has been
lately predicted in $\PT$ materials. We shall elaborate on the ideas
developed by Mostafazadeh~\cite{Mostafazadeh:2013rp} in order to shed
light into this intriguing~question.

The potential $U(x)$ is called reflectionless from the left (right),
if $\coefr^{\ell} (k) = 0$ and $\coefr^{r} (k) \neq 0$ [$\coefr^{r}
(k) = 0$ and $\coefr^{\ell} (k) \neq 0$]. From the explicit matrix
elements in \eqref{eq:ent} and \eqref{eq:SM}, we see that
unidirectional reflectionlessness implies the non-diagonalizability of
both $\matriz{M}$ and $\matriz{S}$. Therefore, the parameters of the
potential for which it becomes reflectionless correspond to
exceptional points of $\matriz{M}$ and
$\matriz{S}$~\cite{Muller:2008gf,Mehri:2008pc}.

The potential is called invisible from the left (right), if it is
reflectionless from left (right) and in addition $\coeft (k)= 1$. We can
easily express the conditions for the unidirectional invisibility
as
\begin{eqnarray}
  M_{12}(k) &\neq & 0 \, , \qquad
  M_{11} (k) =M_{22} (k)  = 1 
  \qquad
  \mathrm{(left \ invisible)} \nonumber \\
  & & \\
  M_{21} (k) & \neq & 0  \, , \qquad
  M_{11} (k) =M_{22} (k)  = 1 
  \qquad
  \mathrm{(right \ invisible)} \nonumber
\end{eqnarray}

Next, we scrutinize the role of $\PT$-symmetry in the invisibility.
For that purpose, we first briefly recall that the parity
transformation ``reflects'' the system with respect to the coordinate
origin, so that $x \mapsto -x $ and the momentum $p \mapsto - p$. The
action on the wave function is
\begin{equation}
  \label{eq:parity}
  \Psi(x) \mapsto  (\Par \Psi) (x) = \Psi (-x)  \,
\end{equation}
On the other hand, the time reversal inverts the sense of time
evolution, so that $ x \mapsto x$, $p \mapsto -p$ and $i \mapsto
-i$. This means that the operator $\T$ implementing such a
transformation is antiunitary and its action~reads
\begin{equation}
  \label{eq:timerev}
  \Psi(x) \mapsto  (\T \Psi)(x) = \Psi^{\ast} (x)  \,
\end{equation}
Consequently, under a combined $\PT$ transformation, we have
\begin{equation}
  \label{eq:moversPT}
  \Psi(x) \mapsto  (\PT \Psi)(x) = \Psi^{\ast} (-x)  \,
\end{equation}

Let us apply this to a general complex scattering potential. The
transfer matrix of the $\PT$-transformed system, we denote by
$\matriz{M}^{(\PT)}$, fulfils
\begin{equation}
  \label{MPT}
  \left (
    \begin{array}{c}
      A_{+}^{\ast} \\
      A_{-}^{\ast}
    \end{array}
  \right ) =
  \matriz{M}^{(\PT)}
  \left (
    \begin{array}{c}
      B_{+}^{\ast} \\
      B_{-}^{\ast}
    \end{array}
  \right ) \,
\end{equation}
Comparing with \eqref{M}, we come to the result
\begin{equation}
  \label{eq:3}
  \matriz{M}^{(\PT)} = (\matriz{M}^{-1})^{\ast} \,
\end{equation}
and,  because $\det \, \matriz{M} = 1$, this means
\begin{equation}
  \label{eq:4}
  M_{11} \stackrel{\PT}{\longmapsto} M_{22}^{\ast} \, ,
  \qquad
  M_{12} \stackrel{\PT}{\longmapsto} - M_{12}^{\ast} \, ,
  \qquad
  M_{21} \stackrel{\PT}{\longmapsto} - M_{21}^{\ast} \, ,
  \qquad
  M_{22} \stackrel{\PT}{\longmapsto} M_{11}^{\ast} \,
\end{equation}
When the system is invariant under this transformation
[$\matriz{M}^{(\PT)} = \matriz{M}$], it must hold
\begin{equation}
  \label{eq:3inv}
  \matriz{M}^{-1} = \matriz{M}^{\ast} \,
\end{equation}
a fact already noticed by Longhi~\cite{Longhi:2010kx} and that can be
also recast as~\cite{Monzon:2013xw}
\begin{equation}
  \label{TMPT}
  \re \left ( \frac{\coefr^{\ell}}{\coeft} \right ) =
  \re \left ( \frac{\coefr^{r}}{\coeft} \right )  = 0 \,
\end{equation}
This can be equivalently restated in the form
\begin{equation}
  \label{eq:coll}
  \rho^{\ell} - \tau = \pm \pi/2 \, ,
  \qquad
  \rho^{r} - \tau = \pm \pi/2 \,
\end{equation}
with $\tau = \arg ( \coeft )$ and $\rho_{\ell, r} = \arg (
\coefr_{\ell, r} )$. Hence, if we look at the complex numbers
$\coefr^{\ell}$, $\coefr^{r}$, and $\coeft$ as phasors,
Equation~(\ref{eq:coll}) tell us that $\coefr^{\ell}$ and $\coefr^{r}$ are
always collinear, while $\coeft$ is simultaneously perpendicular to
them. We draw the attention to the fact that the same expressions have
been derived for lossless symmetric beam
splitters~\cite{Mandel:1995uq}: we have shown that they hold true for
any $\PT$-symmetric structure.

A direct consequence of \eqref{eq:4} is that there are
particular instances of $\PT$-invariant systems that are invisible,
although not every invisible potential is $\PT$ invariant. In this
respect, it is worth stressing, that even ($\Par$-symmetric)
potentials do not support unidirectional invisibility and the same
holds for real ($\T$-symmetric) potentials.

In optics, beam propagation is governed by the paraxial wave equation,
which is equivalent to a Schr\"{o}dinger-like equation,  with the role
of the potential played here by the refractive index. Therefore, a
necessary condition for a complex refractive index to be $\PT$
invariant is that its real part is an even function of $x$,  while the
imaginary component (loss and gain profile) is odd.

\section{Relativistic Variables}

To move ahead, let us construct the Hermitian matrices
\begin{equation}
  \matriz{X} =
  \left (
    \begin{array}{c}
      X_{+}  \\
      X_{-}
    \end{array}
  \right )
  \otimes
  \left (
    \begin{array}{cc}
      X_{+}^{\ast}  &      X_{-}^{\ast}
    \end{array}
  \right ) =
  \left (
    \begin{array}{cc}
      |X_{+}|^2 &  X_{+} X_{-}^{\ast}  \\
      X_{+}^\ast X_{-} & |X_{-}|^2
    \end{array}
  \right ) \,
\end{equation}
where $X_{\pm}$ refers to either $A_{\pm}$ or $B_{\pm}$;  \emph{i.e}.,  the
amplitudes that determine the behavior at each side of the potential.
The matrices $\matriz{X}$ are quite reminiscent of the coherence
matrix in optics or the density matrix in quantum mechanics.

One can  verify that $\matriz{M}$ acts on $\matriz{X}$ by
conjugation
\begin{equation}
  \label{congruence}
  \matriz{X}^{\prime} = \matriz{M} \, \matriz{X} \,
  \matriz{M}^\dagger \,
\end{equation}
The matrix $\matriz{X}^{\prime}$ is associated with the amplitudes $B_{\pm}$
and $\matriz{X}$ with $A_{\pm}$.

Let us consider the set $\sigma^{\mu} = (\openone, \bm{\sigma})$, with
Greek indices running from 0 to 3. The $\sigma^{\mu}$ are the identity and the
standard Pauli matrices, which constitute a basis of the linear space of
$2\times 2$ complex matrices. For the sake of covariance, it is
convenient to define $\tilde{\sigma}^{\mu} \equiv \sigma_{\mu} =
(\openone, - \bm{\sigma})$, so that~\cite{Barut:1977fk}
\begin{equation}
  \label{eq:6}
  \Tr (\tilde{\sigma}^{\mu} \sigma_{\nu})  =
  2 \delta^{\mu}_{\ \nu}  \,
\end{equation}
and $ \delta^{\mu}_{\ \nu}$ is the Kronecker delta.
To any Hermitian matrix $\matriz{X}$ we can associate the coordinates
\begin{equation}
  x^{\mu} =   \textstyle{\frac{1}{2}}  \Tr (\matriz{X}  \tilde{\sigma}^{\mu} )  \,
\end{equation}
The congruence (\ref{congruence}) induces in this way a
transformation
\begin{equation}
  \label{var}
  x^{\prime \, \mu}= \Lambda^{\mu}_{\ \nu} (\matriz{M}) \, x^{\nu}
\end{equation}
where $\Lambda^{\mu}_{\ \nu} (\matriz{M}) $ can be found to be
\begin{equation}
  \label{eq:relation}
  \Lambda^{\mu}_{\ \nu} (\matriz{M} ) =
  \textstyle{\frac{1}{2}} \Tr \left ( \tilde{\sigma}^{\mu} \matriz{M} \sigma_{\nu}
    \matriz{M}^\dagger \right ) \,
\end{equation}
This equation can be solved to obtain $\matriz{M}$ from $\Lambda$.  The
matrices $\matriz{M}$ and $- \matriz{M}$ generate the same $\Lambda$,
so this homomorphism is two-to-one. The variables $x^{\mu}$ are
coordinates in a Minkovskian (1+3)-dimensional space and the action of
the system can be seen as a Lorentz transformation in SO(1, 3).

Having set the general scenario, let us have a closer look at the
transfer matrix corresponding to right invisibility (the left
invisibility can be dealt with in an analogous way); namely,
\begin{equation}
\label{eq:M+}
  \matriz{M} =
  \begin{pmatrix}
    1 & \coefr  \\
    0 & 1
  \end{pmatrix}
\end{equation}
where, for simplicity, we have dropped the superscript from
$\coefr^{r}$, as there is no risk of confusion.  Under the
homomorphism~\eqref{eq:relation} this matrix generates
the Lorentz transformation
\begin{equation}
\label{eq:nullrot}
  \Lambda (\matriz{M}) =
  \begin{pmatrix}
    1+ | \coefr |^2/2 & \re \coefr  & - \im  \coefr  & -| \coefr |^2/2  \\
    \re \coefr        & 1      & 0  & -\re \coefr    \\
    -\im \coefr      & 0      & 1 & \im \coefr          \\
    | \coefr |^2/2 & \re \coefr & - \im \coefr & 1-| \coefr| ^2/2
  \end{pmatrix} \,
\end{equation}
According to Wigner~\cite{Wigner:1939rp}, the little
group is a subgroup of the Lorentz transformations under which a standard vector
$s^{\mu}$ remains invariant. When $s^{\mu}$ is timelike, the little
group is the rotation group SO(3). If $s^{\mu}$ is spacelike,
the little group are the boosts SO(1, 2). In this context, the matrix
(\ref{eq:nullrot}) is an instance of a null rotation;  the little
group when $s^{\mu}$ is a lightlike or null  vector, which is related
to E(2), the symmetry group of the two-dimensional Euclidean
space~\cite{Kim:1986kn}.

If we write (\ref{eq:nullrot}) in the form $ \Lambda (\matriz{M}) =
\exp ( i \matriz{N})$, we can easily work out that
\begin{equation}
  \label{eq:nullrot2}
  \matriz{N}  =
  \begin{pmatrix}
    0 & \re \coefr  & - \im  \coefr  & 0  \\
    \re \coefr        & 0      & 0  & -\re \coefr    \\
    -\im \coefr      & 0      & 0 & \im \coefr          \\
    0 & \re \coefr & - \im \coefr & 0
  \end{pmatrix} \,
\end{equation}
This is a nilpotent matrix and the vectors annihilated by $\matriz{N}$
are invariant by $ \Lambda (\matriz{M})$.  In terms of the Lie algebra
so(1, 3), $\matriz{N}$ can be expressed as
\begin{equation}
  \label{eq:Ngen}
  \matriz{N} = \re R \, (\matriz{K}_{1} + \matriz{J}_{2} )
  - \im R \, (\matriz{K}_{2} + \matriz{J}_{1})  \,
\end{equation}
where $\matriz{K}_{i}$ generate boosts and $\matriz{J}_{i}$ rotations
($i = 1, 2 , 3$)~\cite{Weinberg:2005jl}. Observe that the rapidity of
the boost and the angle of the rotation have the same norm. The matrix
$\matriz{N}$ define a two-parameter Abelian subgroup.

Let us take, for the time being, $\re \coefr=0$, as it happens for
$\PT$-invariant invisibility.  We can express $\matriz{K}_{2} +
\matriz{J}_{1}$ as the differential operator
\begin{equation}
  \label{eq:Killnul}
  \matriz{K}_{2} + \matriz{J}_{1} \mapsto  (x^{2} \partial_{0} + x^{0} \partial_{2})
  + (x^{2} \partial_{3} - x^{3} \partial_{2} ) =  x^{2}  ( \partial_0 + \partial_3 ) +
  (x^{0}-x^{3}) \partial_{2}  \,
\end{equation}
As we can appreciate, the combinations
\begin{equation}
  x^{2} \, , \qquad
  x^{0} - x^{3} \, , \qquad
  (x^{0})^2 - (x^{1})^2- (x^{3})^2 \,
\end{equation}
remain invariant. Suppressing the inessential coordinate $x^{2}$, the
flow lines of the Killing vector \eqref{eq:Killnul} is the
intersection of a null plane, $x^{0} - x^{3} = c_{2 }$ with a
hyperboloid $ (x^{0})^2 - (x^{1})^2- (x^{3})^2= c_3$.  The case
$c_{3}=0$ has the hyperboloid degenerate to a light cone with the
orbits becoming parabolas lying in corresponding null planes.

\section{Hyperbolic Geometry and Invisibility}

Although the relativistic hyperboloid in Minkowski space constitute by
itself a model of hyperbolic geometry (understood in a broad sense, as
the study of spaces with constant negative curvature), it is not the
best suited to display some features.

Let us consider the customary tridimensional hyperbolic
space $\mathbb{H}^{3}$, defined in terms of the upper half-space
$\{ (x, y , z ) \in \mathbb{R}^{3} | z > 0 \}$,
equipped with the metric~\cite{Iversen:1992qc}
\begin{equation}
  \label{eq:hypmet}
  ds^{2} =  \frac{\sqrt{dx^{2} + dy^{2} +dz^{2}}}{z} \,
\end{equation}
The geodesics are the semicircles in $\mathbb{H}^{3}$ orthogonal to
the plane $z = 0$.

We can think of the plane $z=0$ in $\mathbb{R}^{3}$ as the complex
plane $\mathbb{C}$ with the natural identification $(x, y, z) \mapsto
w= x + i y $. We need to add the point at infinity, so that
$\hat{\mathbb{C}} = \mathbb{C} \cup \infty$, which is usually referred
to as the Riemann sphere and identify $\hat{\mathbb{C}}$ as the
boundary of $\mathbb{H}^{3}$.

Every matrix $\matriz{M}$ in SL(2, $\mathbb{C}$) induces a natural
mapping in $\mathbb{C}$ via M\"{o}bius (or bilinear)
\mbox{transformations~\cite{Ratcliffe:2006sy}}
\begin{equation}
  \label{accion}
  w^{\prime} =   \frac {M_{11} w + M_{12}}{M_{21} w + M_{22} } \,
\end{equation}
Note that any matrix obtained by multiplying $\matriz{M}$ by a complex
scalar $\lambda$ gives the same transformation, so a M\"{o}bius
transformation determines its matrix only up to scalar multiples. In
other words, we need to quotient out SL(2, $\mathbb{C})$ by its center
$\{ \openone, - \openone \}$: the resulting quotient group is known as
the projective linear group and is usually denoted PSL(2,
$\mathbb{C}$).

Observe that we can break down the action \eqref{accion} into a
composition of maps of the form
\begin{equation}
  \label{eq:elemap}
  w \mapsto w + \lambda \, ,
  \qquad
  w \mapsto \lambda w \, ,
  \qquad
  w \mapsto - 1/w \,
\end{equation}
with $\lambda \in \mathbb{C}$. Then we can extend the M\"{o}bius
transformations to all $\mathbb{H}^{3}$ as follows:
\begin{equation}
  \label{eq:extmob}
  (w, z) \mapsto (w + \lambda, z) \, ,
  \qquad
  (w, z)  \mapsto (\lambda w, |\lambda| z) \, ,
  \qquad
  (w , z ) \mapsto \left ( - \frac{w^{\ast}}{|w^{2} | + z^{2}},
 \frac{z}{|w^{2} | + z^{2}} \right ) \,
\end{equation}
The expressions above come from decomposing the action on
$\hat{\mathbb{C}}$ of each of the elements of PSL(2, $\mathbb{C}$) in
question into two inversions (reflections) in circles in
$\hat{\mathbb{C}}$. Each such inversion has a unique extension to
$\mathbb{H}_{3}$ as an inversion in the hemisphere spanned by the
circle and composing appropriate pairs of inversions gives us these
formulas.

In fact, one can show that PSL(2, $\mathbb{C}$) preserves the metric
on $\mathbb{H}_{3}$. Moreover every isometry of $\mathbb{H}_{3}$ can
be seen to be the extension of a conformal map of $\hat{\mathbb{C}}$
to itself, since it must send hemispheres orthogonal to
$\hat{\mathbb{C}}$ to hemispheres orthogonal to $\hat{\mathbb{C}}$,
hence circles in $\hat{\mathbb{C}}$ to circles in
$\hat{\mathbb{C}}$. Thus all orientation-preserving isometries of
$\mathbb{H}_{3}$ are given by elements of PSL(2, $\mathbb{C}$) acting
as above.

In the classification of these isometries the notion of fixed points
is of utmost importance. These points are defined by the condition
$w^{\prime} = w $ in \eqref{accion}, whose solutions are
\begin{equation}
  w_{f} = \frac{(M_{11}-M_{22}) \pm \sqrt{ [ \Tr ( \matriz{M} ) ]^2 -4}}{2M_{21}} \,
\end{equation}
So, they are determined by the trace of $\matriz{M}$.  When the trace
is a real number, the induced M\"{o}bius transformations are called
elliptic, hyperbolic, or parabolic, according $[\Tr ( \matriz{M} )]
^2$ is lesser than, greater than, or equal to 4, respectively. The
canonical representatives of those matrices are~\cite{Anderson:1999bs}
\begin{equation}
  \underbrace{\left (
      \begin{array}{cc}
        e^{i \theta/2} &  0 \\
        0 &  e^{-i \theta/2}
      \end{array}
    \right )}_{\mbox{\footnotesize elliptic}} \, ,
  \quad
  \underbrace{\left (
      \begin{array}{cc}
        e^{\xi/2} &  0 \\
        0 &  e^{-\xi/2}
      \end{array}
    \right )}_{\mbox{\footnotesize hyperbolic}}  \, ,
  \quad
  \underbrace{\left (
      \begin{array}{cc}
        1 &  \lambda \\
        0 &  1
      \end{array}
    \right )}_{\mbox{\footnotesize parabolic}} \,
\end{equation}
while the induced geometrical actions are
\begin{equation}
  \label{eq:1}
  w^{\prime} = w e^{i\theta} \, ,
  \qquad
  w^{\prime} = w e^{\xi} \,  ,
  \qquad
  w^{\prime}  = w + \lambda  \,
\end{equation}
that is, a rotation of angle $\theta$ (so fixes the axis $z$), a
squeezing of parameter $\xi$ (it has two fixed points in
$\hat{\mathbb{C}}$, no fixed points in $\mathbb{H}_{3}$, and every
hyperplane in $\mathbb{H}_{3}$ that contains the geodesic joining the
two fixed points in $\hat{\mathbb{C}}$ is invariant); and a parallel
displacement of magnitude $\lambda$, respectively. We emphasize that
this later action is the only one without Euclidean analogy. Indeed,
in view of \eqref{eq:M+}, this is precisely the action associated to
an invisible scatterer. The far-reaching consequences of this
geometrical interpretation will be developed elsewhere.

\section{Concluding Remarks}

We have studied unidirectional invisibility by a complex scattering
potential, which is characterized by a set of $\PT$ invariant
equations. Consequently, the $\PT$-symmetric invisible configurations
are quite special, for they possess the same symmetry as the
equations.

We have shown how to cast this phenomenon in term of space-time
variables, having in this way a relativistic presentation of
invisibility as the set of null rotations. By resorting to elementary
notions of hyperbolic geometry, we have interpreted in a natural way
the action of the transfer matrix in this case as a parallel
displacement.

We think that our results are yet another example of the advantages of
these geometrical methods: we have devised a geometrical tool to
analyze invisibility in quite a concise way that, in addition, can be
closely related to other fields of physics.

\acknowledgements{Acknowledgements}

We acknowledge illuminating discussions with Antonio F. Costa, Jos\'e
F. Cari\~{n}ena and Jos\'e Mar\'{\i}a Montesinos.  Financial support
from the Spanish Research Agency (Grant FIS2011-26786) is gratefully
acknowledged.

\section*{\noindent Author Contributions}
\vspace {12pt}

Both authors contributed equally to all aspects of preparing this
manuscript. 


\section*{\noindent Conflicts of Interest}
\vspace {12pt}

The authors declare no conflicts of interest.


\begin{thebibliography}{----}
\providecommand{\natexlab}[1]{#1}

\bibitem[Bender and Boettcher(1998)]{Bender:1998fk}
Bender, C.M.; Boettcher, S.
\newblock Real spectra in non-Hermitian Hamiltonians having PT symmetry.
\newblock {\em Phys. Rev. Lett.} {\bf 1998},
\newblock {\em 80},~5243--5246.

\bibitem[Bender \em{et~al.}(1999)Bender, Boettcher, and
  Meisinger]{Bender:1999vn}
Bender, C.M.; Boettcher, S.; Meisinger, P.N.
\newblock PT-symmetric quantum mechanics.
\newblock {\em J. Math. Phys.} {\bf 1999},
\newblock {\em 40},~2201--2229.

\bibitem[Bender \em{et~al.}(2002)Bender, Brody, and Jones]{Bender:2002uq}
Bender, C.M.; Brody, D.C.; Jones, H.F.
\newblock Complex extension of quantum mechanics.
\newblock {\em Phys. Rev. Lett.} {\bf 2002},
\newblock {\em 89},~270401.

\bibitem[Bender \em{et~al.}(2003)Bender, Brody, and Jones]{Bender:2003zr}
Bender, C.M.; Brody, D.C.; Jones, H.F.
\newblock Must a Hamiltonian be Hermitian?
\newblock {\em Am. J. Phys.} {\bf 2003},
\newblock {\em 71},~1095--1102.

\bibitem[Bender(2007)]{Bender:2007ve}
Bender, C.M.
\newblock Making sense of non-Hermitian Hamiltonians.
\newblock {\em Rep. Prog. Phys.} {\bf 2007},
\newblock {\em 70}, 947--1018.

\bibitem[Bender and Mannheim(2010)]{Bender:2010ly}
Bender, C.M.; Mannheim, P.D.
\newblock PT symmetry and necessary and sufficient conditions for the reality
  of energy eigenvalues.
\newblock {\em Phys. Lett. A} {\bf 2010},
\newblock {\em 374},~1616--1620.

\bibitem[Assis(2010)]{Assis:2010fk}
Assis, P.
\newblock {\em Non-{H}ermitian {H}amiltonians in {F}ield {T}heory:
  {PT}-symmetry and {A}pplications}; VDM: Saarbr{\"u}cken, Germany,
\newblock  2010.

\bibitem[Moiseyev(2011)]{Moiseyev:2011qy}
Moiseyev, N.
\newblock {\em Non-{H}ermitian {Q}uantum {M}echanics}; Cambridge {U}niversity
  {P}ress: Cambridge, UK,
\newblock  2011.

\bibitem[El-Ganainy \em{et~al.}(2007)El-Ganainy, Makris, Christodoulides, and
  Musslimani]{El-Ganainy:2007mb}
El-Ganainy, R.; Makris, K.G.; Christodoulides, D.N.; Musslimani, Z.H.
\newblock Theory of coupled optical PT-symmetric structures.
\newblock {\em Opt. Lett.} {\bf 2007},
\newblock {\em 32},~2632--2634.

\bibitem[Bendix \em{et~al.}(2009)Bendix, Fleischmann, Kottos, and
  Shapiro]{Bendix:2009gb}
Bendix, O.; Fleischmann, R.; Kottos, T.; Shapiro, B.
\newblock Exponentially fragile PT symmetry in lattices with localized
  eigenmodes.
\newblock {\em Phys. Rev. Lett.} {\bf 2009},
\newblock {\em 103},~030402.

\bibitem[Ruter \em{et~al.}(2010)Ruter, Makris, El-Ganainy, Christodoulides,
  Segev, and Kip]{Ruter:2010ss}
Ruter, C.E.; Makris, K.G.; El-Ganainy, R.; Christodoulides, D.N.; Segev, M.;
  Kip, D.
\newblock Observation of parity-time symmetry in optics.
\newblock {\em Nat. Phys.} {\bf 2010},
\newblock {\em 6},~192--195.

\bibitem[Makris \em{et~al.}(2008)Makris, El-Ganainy, Christodoulides, and
  Musslimani]{Makris:2008jl}
Makris, K.G.; El-Ganainy, R.; Christodoulides, D.N.; Musslimani, Z.H.
\newblock Beam dynamics in PT symmetric optical lattices.
\newblock {\em Phys. Rev. Lett.} {\bf 2008},
\newblock {\em 100},~103904.

\bibitem[Longhi(2009)]{Longhi:2009uq}
Longhi, S.
\newblock Bloch oscillations in complex crystals with PT symmetry.
\newblock {\em Phys. Rev. Lett.} {\bf 2009},
\newblock {\em 103},~123601.

\bibitem[Sukhorukov \em{et~al.}(2010)Sukhorukov, Xu, and
  Kivshar]{Sukhorukov:2010ys}
Sukhorukov, A.A.; Xu, Z.; Kivshar, Y.S.
\newblock Nonlinear suppression of time reversals in PT-symmetric optical
  couplers.
\newblock {\em Phys. Rev. A} {\bf 2010},
\newblock {\em 82},~043818.

\bibitem[Ahmed \em{et~al.}(2005)Ahmed, Bender, and Berry]{Ahmed:2005fk}
Ahmed, Z.; Bender, C.M.; Berry, M.V.
\newblock Reflectionless potentials and PT symmetry.
\newblock {\em J. Phys. A} {\bf 2005},
\newblock {\em 38},~L627--L630.

\bibitem[Lin \em{et~al.}(2011)Lin, Ramezani, Eichelkraut, Kottos, Cao, and
  Christodoulides]{Lin:2011zr}
Lin, Z.; Ramezani, H.; Eichelkraut, T.; Kottos, T.; Cao, H.; Christodoulides,
  D.N.
\newblock Unidirectional invisibility dnduced by PT-symmetric periodic
  structures.
\newblock {\em Phys. Rev. Lett.} {\bf 2011},
\newblock {\em 106},~213901.

\bibitem[Longhi(2011)]{Longhi:2011ve}
Longhi, S.
\newblock Invisibility in PT-symmetric complex crystals.
\newblock {\em J. Phys. A} {\bf 2011},
\newblock {\em 44},~485302.

\bibitem[S{\'a}nchez-Soto \em{et~al.}(2012)S{\'a}nchez-Soto, Monz{\'o}n,
  Barriuso, and Cari{\~n}ena]{Sanchez-Soto:2012bh}
S{\'a}nchez-Soto, L.L.; Monz{\'o}n, J.J.; Barriuso, A.G.; Cari{\~n}ena, J.
\newblock The transfer matrix: A geometrical perspective.
\newblock {\em Phys. Rep.} {\bf 2012},
\newblock {\em 513},~191--227.

\bibitem[Monz\'{o}n and S\'{a}nchez-Soto(1999)]{Monzon:1999eh}
Monz\'{o}n, J.J.; S\'{a}nchez-Soto, L.L.
\newblock Lossles multilayers and Lorentz transformations: More than an
  analogy.
\newblock {\em Opt. Commun.} {\bf 1999},
\newblock {\em 162},~1--6.

\bibitem[Monz{\'{o}}n and S\'{a}nchez-Soto(1999)]{Monzon:1999fk}
Monz{\'{o}}n, J.J.; S\'{a}nchez-Soto, L.L.
\newblock Fullly relativisticlike formulation of multilayer optics.
\newblock {\em J. Opt. Soc. Am. A} {\bf 1999},
\newblock {\em 16},~2013--2018.

\bibitem[J.~J.~Monz{\'{o}}n and S{\'{a}}nchez-Soto(2001)]{Monzon:2001b}
J.~J.~Monz{\'{o}}n, T.Y.; S{\'{a}}nchez-Soto, L.L.
\newblock Basic factorization for multilayers.
\newblock {\em Opt. Lett.} {\bf 2001},
\newblock {\em 26},~370--372.

\bibitem[Yonte \em{et~al.}(2002)Yonte, Monz\'on, S\'anchez-Soto,
  Cari{\~{n}}ena, and L\'opez-Lacasta]{Yonte:2002}
Yonte, T.; Monz\'on, J.J.; S\'anchez-Soto, L.L.; Cari{\~{n}}ena, J.F.;
  L\'opez-Lacasta, C.
\newblock Understanding multilayers from a geometrical viewpoint.
\newblock {\em J. Opt. Soc. Am. A} {\bf 2002},
\newblock {\em 19},~603--609.

\bibitem[Monz{\'{o}}n \em{et~al.}(2002)Monz{\'{o}}n, Yonte, S{\'a}nchez-Soto,
  and Cari{\~n}ena]{Monzon:2002mz}
Monz{\'{o}}n, J.J.; Yonte, T.; S{\'a}nchez-Soto, L.L.; Cari{\~n}ena, J.F.
\newblock Geometrical setting for the classification of multilayers.
\newblock {\em J. Opt. Soc. Am. A} {\bf 2002},
\newblock {\em 19},~985--991.

\bibitem[Barriuso \em{et~al.}(2003)Barriuso, Monz{\'{o}}n, and
  S{\'{a}}nchez{-S}oto]{Barriuso:2003}
Barriuso, A.G.; Monz{\'{o}}n, J.J.; S{\'{a}}nchez{-S}oto, L.L.
\newblock General unit{-}disk representation for periodic multilayers.
\newblock {\em Opt. Lett.} {\bf 2003},
\newblock {\em 28},~1501--1503.

\bibitem[Barriuso \em{et~al.}(2004)Barriuso, Monz{\'{o}}n,
  S{\'{a}}nchez{-}{S}oto, and Cari{\~{n}}ena]{Barriuso:2004}
Barriuso, A.G.; Monz{\'{o}}n, J.J.; S{\'{a}}nchez{-}{S}oto, L.L.;
  Cari{\~{n}}ena, J.F.
\newblock Vectorlike representation of multilayers.
\newblock {\em J. Opt. Soc. Am. A} {\bf 2004},
\newblock {\em 21},~2386--2391.

\bibitem[Barriuso \em{et~al.}(2009)Barriuso, Monz{\'{o}}n, S\'{a}nchez-Soto,
  and Costa]{Barriuso:2009}
Barriuso, A.G.; Monz{\'{o}}n, J.J.; S\'{a}nchez-Soto, L.L.; Costa, A.F.
\newblock Escher-like quasiperiodic heterostructures.
\newblock {\em J. Phys. A} {\bf 2009},
\newblock {\em 42},~192002.

\bibitem[Muga \em{et~al.}(2004)Muga, Palao, Navarro, and
  Egusquiza]{Muga:2004ys}
Muga, J.G.; Palao, J.P.; Navarro, B.; Egusquiza, I.L.
\newblock Complex absorbing potentials.
\newblock {\em Phys. Rep.} {\bf 2004},
\newblock {\em 395},~357--426.

\bibitem[Levai and Znojil(2000)]{Levai:2000kx}
Levai, G.; Znojil, M.
\newblock Systematic search for PT-symmetric potentials with real spectra.
\newblock {\em J. Phys. A} {\bf 2000},
\newblock {\em 33},~7165--7180.

\bibitem[Ahmed(2001{\natexlab{a}})]{Ahmed:2001ys}
Ahmed, Z.
\newblock Schr{\"o}dinger transmission through one-dimensional complex
  potentials.
\newblock {\em Phys. Rev. A} {\bf 2001},
\newblock {\em 64},~042716.

\bibitem[Ahmed(2001{\natexlab{b}})]{Ahmed:2001dz}
Ahmed, Z.
\newblock Energy band structure due to a complex, periodic, PT-invariant
  potential.
\newblock {\em Phys. Lett. A} {\bf 2001},
\newblock {\em 286},~231--235.

\bibitem[Mostafazadeh(2009)]{Mostafazadeh:2009ve}
Mostafazadeh, A.
\newblock Spectral singularities of complex scattering potentials and infinite
  reflection and transmission coefficients at real energies.
\newblock {\em Phys. Rev. Lett.} {\bf 2009},
\newblock {\em 102},~220402.

\bibitem[Cannata \em{et~al.}(2007)Cannata, Dedonder, and
  Ventura]{Cannata:2007kx}
Cannata, F.; Dedonder, J.P.; Ventura, A.
\newblock Scattering in PT-symmetric quantum mechanics.
\newblock {\em Ann. Phys.} {\bf 2007},
\newblock {\em 322},~397--433.

\bibitem[Chong \em{et~al.}(2011)Chong, Ge, and Stone]{Chong:2011vn}
Chong, Y.D.; Ge, L.; Stone, A.D.
\newblock PT-symmetry breaking and laser-absorber modes in optical scattering
  systems.
\newblock {\em Phys. Rev. Lett.} {\bf 2011},
\newblock {\em 106},~093902.

\bibitem[Ahmed(2012)]{Ahmed:2012qf}
Ahmed, Z.
\newblock New features of scattering from a one-dimensional non-Hermitian
  (complex) potential.
\newblock {\em J. Phys. A} {\bf 2012},
\newblock {\em 45},~032004.

\bibitem[Boonserm and Visser(2010)]{Boonserm:2010fk}
Boonserm, P.; Visser, M.
\newblock One dimensional scattering problems: A pedagogical presentation of
  the relationship between reflection and transmission amplitudes.
\newblock {\em Thai J. Math.} {\bf 2010},
\newblock {\em 8},~83--97.

\bibitem[Mostafazadeh and Mehri-Dehnavi(2009)]{Mostafazadeh:2009yo}
Mostafazadeh, A.; Mehri-Dehnavi, H.
\newblock Spectral singularities, biorthonormal systems and a \linebreak two-parameter
  family of complex point interactions.
\newblock {\em J. Phys. A} {\bf 2009},
\newblock {\em 42},~125303.

\bibitem[Aktosun(1992)]{Aktosun:1992kx}
Aktosun, T.
\newblock A factorization of the scattering matrix for the {S}chr{\"o}dinger
  equation and for the wave equation in one dimension.
\newblock {\em J. Math. Phys.} {\bf 1992},
\newblock {\em 33},~3865--3869.

\bibitem[Aktosun \em{et~al.}(1996)Aktosun, Klaus, and van~der
  Mee]{Aktosun:1996fj}
Aktosun, T.; Klaus, M.; van~der Mee, C.
\newblock Factorization of scattering matrices due to partitioning of
  potentials in one{-}dimensional {S}chr{\"o}dinger{-}type equations.
\newblock {\em J. Math. Phys.} {\bf 1996},
\newblock {\em 37},~5897--5915.

\bibitem[Marchenko(1986)]{Marchenko:1986fk}
Marchenko, V.A.
\newblock {\em Sturm-{L}iouville {O}perators and {T}heir {A}pplications}; AMS
  Chelsea: Providence, RI, USA,
\newblock  1986.

\bibitem[Tunca and Bairamov(1999)]{Tunca:1999vl}
Tunca, G.; Bairamov, E.
\newblock Discrete spectrum and principal functions of non-selfadjoint
  differential operator.
\newblock {\em Czech J. Math.} {\bf 1999},
\newblock {\em 49},~689--700.

\bibitem[Naimark(1960)]{Naimark:1960bv}
Naimark, M.A.
\newblock Investigation of the spectrum and the expansion in eigenfunctions of
  a non-selfadjoint operator of the second order on a semi-axis.
\newblock {\em AMS Transl.} {\bf 1960},
\newblock {\em 16},~103--193.

\bibitem[Pavlov(1967)]{Pavlov:1967kb}
Pavlov, B.S.
\newblock The nonself-adjoint Schr\"{o}dinger operators.
\newblock {\em Topics Math. Phys.} {\bf 1967},
\newblock {\em 1},~87--114.

\bibitem[Naimark(1968)]{Naimark:1968qr}
Naimark, M.A.
\newblock {\em Linear Differential Operators: Part II}; Ungar: New York, NY, USA,
\newblock  1968.

\bibitem[Samsonov(2005)]{Samsonov:2005qq}
Samsonov, B.F.
\newblock {SUSY} transformations between diagonalizable and non-diagonalizable
  \linebreak Hamiltonians.
\newblock {\em J. Phys. A} {\bf 2005},
\newblock {\em 38},~L397--L403.

\bibitem[Andrianov \em{et~al.}(2010)Andrianov, Cannata, and
  Sokolov]{Andrianov:2010ts}
Andrianov, A.A.; Cannata, F.; Sokolov, A.V.
\newblock Spectral singularities for non-Hermitian one-dimensional
  Hamiltonians: Puzzles with resolution of identity.
\newblock {\em J. Math. Phys.} {\bf 2010},
\newblock {\em 51},~052104.

\bibitem[Chaos-Cador and Garc{\'\i}a-Calder{\'o}n(2013)]{Chaos-Cador:2013rf}
Chaos-Cador, L.; Garc{\'\i}a-Calder{\'o}n, G.
\newblock Resonant states for complex potentials and spectral singularities.
\newblock {\em Phys. Rev. A} {\bf 2013},
\newblock {\em 87},~042114.

\bibitem[Schomerus(2010)]{Schomerus:2010dk}
Schomerus, H.
\newblock Quantum noise and self-sustained radiation of PT-symmetric systems.
\newblock {\em Phys. Rev. Lett.} {\bf 2010},
\newblock {\em 104},~233601.

\bibitem[Longhi(2010)]{Longhi:2010kx}
Longhi, S.
\newblock PT-symmetric laser absorber.
\newblock {\em Phys. Rev. A} {\bf 2010},
\newblock {\em 82},~031801.

\bibitem[Mostafazadeh(2013{\natexlab{a}})]{Mostafazadeh:2013uf}
Mostafazadeh, A.
\newblock Nonlinear spectral singularities of a complex barrier potential and
  the lasing threshold condition.
\newblock {\em Phys. Rev. A} {\bf 2013},
\newblock {\em 87},~063838.

\bibitem[Mostafazadeh(2013{\natexlab{b}})]{Mostafazadeh:2013rp}
Mostafazadeh, A.
\newblock Invisibility and PT symmetry.
\newblock {\em Phys. Rev. A} {\bf 2013},
\newblock {\em 87},~012103.

\bibitem[M{\"u}ller and Rotter(2008)]{Muller:2008gf}
M{\"u}ller, M.; Rotter, I.
\newblock Exceptional points in open quantum systems.
\newblock {\em J. Phys. A} {\bf 2008},
\newblock {\em 41},~244018.

\bibitem[Mehri-Dehnavi and Mostafazadeh(2008)]{Mehri:2008pc}
Mehri-Dehnavi, H.; Mostafazadeh, A.
\newblock Geometric phase for non-Hermitian Hamiltonians and its holonomy
  interpretation.
\newblock {\em J. Math. Phys.} {\bf 2008},
\newblock {\em 49},~082105.

\bibitem[Monz{\'o}n \em{et~al.}(2013)Monz{\'o}n, Barriuso, Montesinos-Amilibia,
  and S{\'a}nchez-Soto]{Monzon:2013xw}
Monz{\'o}n, J.J.; Barriuso, A.G.; Montesinos-Amilibia, J.M.; S{\'a}nchez-Soto,
  L.L.
\newblock Geometrical aspects of PT-invariant transfer matrices.
\newblock {\em Phys. Rev. A} {\bf 2013},
\newblock {\em 87},~012111.

\bibitem[Mandel and Wolf(1995)]{Mandel:1995uq}
Mandel, L.; Wolf, E.
\newblock {\em Optical Coherence and Quantum Optics}; Cambridge University
  Press: Cambridge, UK,
\newblock  1995. Sect. 12.12.

\bibitem[Barut and R{\c{a}}czka(1977)]{Barut:1977fk}
Barut, A.O.; R{\c{a}}czka, R.
\newblock {\em Theory of {G}roup {R}epresentations and {A}pplications}; PWN:
  Warszaw, Poland,
\newblock  1977; Sect. 17.2.

\bibitem[Wigner(1939)]{Wigner:1939rp}
Wigner, E.
\newblock On unitary representations of the inhomogeneous {L}orentz group.
\newblock {\em Ann. Math.} {\bf 1939},
\newblock {\em 40},~149--204.

\bibitem[Kim and Noz(1986)]{Kim:1986kn}
Kim, Y.S.; Noz, M.E.
\newblock {\em Theory and {A}pplications of the {P}oincar{\'{e}} {G}roup};
  Reidel: Dordrecht, The Netherlands,
\newblock  1986.

\bibitem[Weinberg(2005)]{Weinberg:2005jl}
Weinberg, S.
\newblock {\em The {Q}uantum {T}heory of {F}ields};  Cambridge
  {U}niversity {P}ress: Cambridge, UK,
\newblock  2005; Volume~1.

\bibitem[Iversen(1992)]{Iversen:1992qc}
Iversen, B.
\newblock {\em Hyperbolic {G}eometry}; Cambridge {U}niversity {P}ress:
  Cambridge, UK,
\newblock  1992; \mbox{Chapter VIII}.

\bibitem[Ratcliffe(2006)]{Ratcliffe:2006sy}
Ratcliffe, J.G.
\newblock {\em Foundations of Hyperbolic Manifolds}; Springer: Berlin, Germany,
\newblock  2006; Sect. 4.3.

\bibitem[Anderson(1999)]{Anderson:1999bs}
Anderson, J.W.
\newblock {\em Hyperbolic {G}eometry}; Springer: New York, NY, USA,
\newblock  1999; Chapter 3.

\end{thebibliography}

\end{document}